\documentclass[superscriptaddress,prd,nofootinbib,twocolumn]{revtex4-2}
\usepackage{graphicx}
\usepackage{amsmath}
\usepackage{amssymb}
\usepackage{amsfonts}
\usepackage{color}
\usepackage{xcolor}
\usepackage{dsfont}
\usepackage{tikz}
\usepackage{ragged2e}
\usetikzlibrary{calc,decorations.markings}
\usepackage[colorlinks=true,linkcolor=blue,citecolor=blue,urlcolor=blue]{hyperref}
\usepackage{physics}
\usepackage{soul}
\usepackage{float}
\usepackage{subfig}
\newcommand{\xx}{\boldsymbol{x}}
\newcommand{\ri}{\boldsymbol{r}_i}
\newcommand{\br}{\boldsymbol{r}}

\newcommand{\bey}{\begin{eqnarray}}
\newcommand{\eey}{\end{eqnarray}}

\begin{document}

\title{Gravity-mediated decoherence}

\author{Dimitris Moustos}
\email{dmoustos@upatras.gr}
\affiliation{Department of Physics, University of Patras, 26504 Patras, Greece}
\author{Charis Anastopoulos}
\email{anastop@upatras.gr}
\affiliation{Department of Physics, University of Patras, 26504 Patras, Greece}
\date{\today}

\begin{abstract}
A small quantum system within the gravitational field of a massive body will be entangled with the quantum degrees of freedom of the latter. Hence, the massive body acts as an environment, and it induces non-unitary dynamics, noise, and decoherence to the quantum system. It is impossible to shield systems on Earth from this gravity-mediated decoherence, which could severely affect all experiments with macroscopic quantum systems. We undertake a first-principles analysis of this effect,  by deriving the corresponding open system dynamics. We find that near-future quantum experiments are not affected, but there is a strong decoherence effect at the human scale. The decoherence time for a  superposition of two localized states of a human with an one meter separation   is of the order of one second.
\end{abstract}

\maketitle


\section{Introduction}\
Contrasting the existence of many hypotheses or theories about a quantum theory of gravity, the experimental information about the interplay of gravity and quantum theory is surprisingly sparse, even  in the non-relativistic, weak-field regime. The famous Colella-Overhauser-Werner (COW) experiment \cite{COW} established that the  effects of a background gravitational field on non-relativistic particles are accounted by the addition of a potential term in the Hamiltonian operator. Later experiments on neutrons bouncing off a horizontal mirror \cite{nbounce}  demonstrated  the existence of bound states due to the gravitational field.

 There is as yet no experimental test of the gravitational interaction between two different quantum matter distributions. In classical physics, gravity is universal, that is, it affects all bodies. It is always attractive, so it is impossible to shield any body from its effects. Furthermore, in  
  the non-relativistic weak-field limit, gravity is non-dynamical. It is described solely by the gravitational potential, which is completely slaved to the mass density through Poisson's equation. 

Taking the gravitational potential to be slaved to the mass density also for quantum systems is perhaps the most conservative assumption about the relation of gravity and quantum theory in the weak-field regime. Nonetheless, it has profound implications.  It implies the possibility of gravitational Schr\"odinger-cat states \cite{AnHu15}, that is, of measurable superpositions of the gravitational force. It also implies that the gravitational interaction may induce quantum correlations, such as entanglement \cite{Bose17, Vedral17, BCM18, AnHu20}, that are experimentally accessible. 

This means that a small quantum particle within the gravitational field of a massive body---for example, the Earth--- is entangled with the quantum degrees of freedom of the latter. Hence, when considering the reduced dynamics of the small particle, the massive body plays the role of an environment, and it leads to non-unitary dynamics, noise, and decoherence. 

In this paper, we undertake a first-principles  analysis of open system dynamics and decoherence for a particle in the gravitational field of a heavy body. We call this type of decoherence ``gravity mediated", rather than ``gravitational", because it originates from the quantum fluctuations of matter. Gravity plays the role of the transmission channel for those fluctuations, unlike gravitational decoherence models, in which  the source of  decoherence is the gravitational field itself \cite{gravdec1, gravdec2}.

The main motivation for our analysis is that gravity-mediated decoherence affects any system inside the gravitational field of the Earth. It sets an upper limit to the size of any macroscopic superpositions that can be created in a  terrestrial laboratory. The existence of such a limit is unavoidable, but its value cannot be estimated with simple arguments: a detailed analysis and modelling is necessary. There is a good {\em a priori} possibility that this limit would affect many proposed experiments that involve macroscopic superpositions. This includes tests of dynamical reduction theories \cite{reduc,reduc2}, tests of fundamental/gravitational decoherence \cite{gravdec1, gravdec2}, and generation of gravity-induced effects, such as entanglement \cite{Bose17, Vedral17, BCM18, AnHu20}. It is therefore essential to have a quantitative estimate of the strength of gravity-mediated decoherence.

We undertake a first-principles analysis of the effect, and we evaluate the resulting  decoherence rate. For Earth, this rate is small; it will not affect currently proposed experiments on macroscopic quantum systems. However, it is not negligible. Gravity-mediated decoherence effects become significant when quantum superpositions reach the scale of humans (or of cats). The decoherence time for a  superposition of two localized states with mean separation of one meter for a human is of the order of one second. Macroscopic superpositions of significantly heavier bodies are not possible on the surface of the Earth. However, the decoherence rate drops as $d^{-3}$, where $d$ is a quantum particle's distance from the surface of the Earth. The limits set by gravity-mediated decoherence are not fundamental, but practical.

To obtain these results,  we derived the open system dynamics for a particle inside the gravitational field generated by a large body. This dynamics is well approximated by a quantum Brownian motion (QBM) model \cite{Feyn:Vern,CaLe,Grabert,HPZ}, which is exactly solvable. 

Gravity-mediated decoherence affects quantum systems in the vicinity of all massive bodies, except for black holes. The latter are vacuum solutions to Einstein's equation, and they do not contain any quantum matter to generate decoherence. The model constructed here applies to any gravitating system for which the Newtonian description of gravity is a good approximation. For example, it applies to quantum particles in the vicinity of compact stars, such as white dwarfs.

The structure of this paper is the following. In Sec. \ref{setup}, we develop the dynamics of our model, and we show that it is mathematically equivalent to QBM. In Sec. \ref{sec:master}, we present the master equation and derive a general formula for the decoherence rate. In Sec. \ref{sec:noise}, we make some simplifying approximations, in order to perform an explicit calculation of the decoherence rate. In Sec. \ref{sec:concl}, we summarize and discuss our results.

\section{Setup}\label{setup}

\begin{figure}[t]
    \centering
    \includegraphics[scale=0.6]{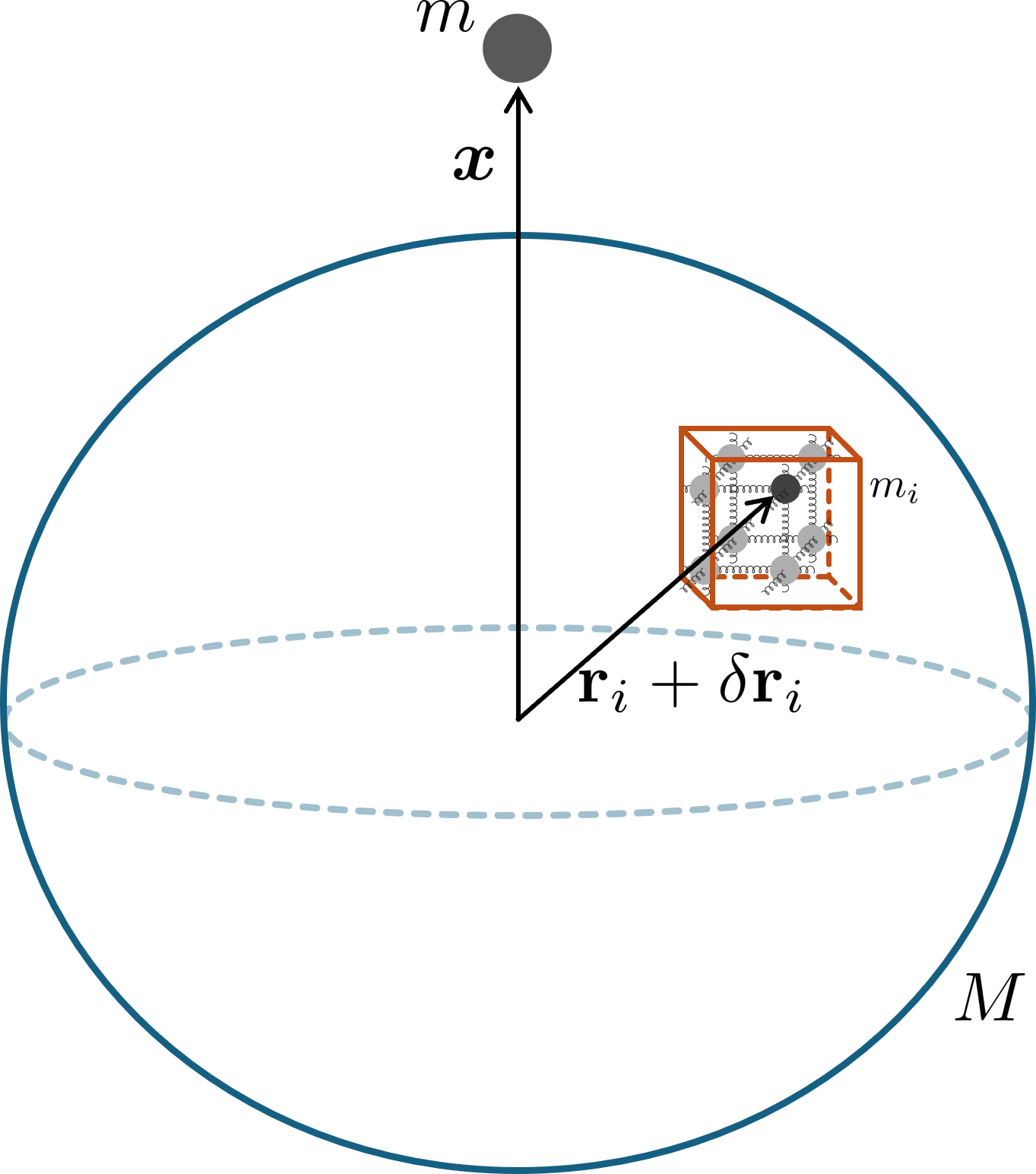}
    \caption{A particle of mass $m$ lies within the gravitational field of a massive body of a total mass $M$. The large body can be described as a collection of cubic cells, each localized around a point $\br_i$, and representing a crystal lattice of oscillators.
    }
    \label{fig:setup}
\end{figure}

We consider a particle of mass $m$ under a potential $V({\xx})$ localized at $\xx$, and interacting gravitationally with a spherically symmetric mass distribution of total mass $M$ (see Fig. \ref{fig:setup}). We assume that the massive body is composed of a finite collection of uncoupled harmonic oscillators of masses $m_i$ and frequencies $\omega_i$, each located at $\ri+\delta \ri$, where the term  $\delta\ri$ describes small displacements from each oscillator's equilibrium position. The  Hamiltonian of the combined system reads
\begin{align}
H_{\text{tot}}=H_{\text{m}}+H_{\text{M}}+H_{\text{int}}, 
\end{align}
where
\begin{align}
    H_{\text{m}}&=\frac{\boldsymbol{p}^2}{2m}+ V({\xx}),\\
    H_{\text{M}}&=\sum_i\left(\frac{\boldsymbol{p}_i^2}{2m_i}+\frac{1}{2}m_i\omega_i^2\delta\ri^2\right),
\end{align}
are respectively the free Hamiltonians of the particle and the harmonic oscillators, and $H_{\text{int}}$ is the interaction term between the small particle of mass $m$ and each of the harmonic oscillators that compose the heavy body. The interaction Hamiltonian is described by the Newtonian gravitational potential
\begin{equation}
    H_{\text{int}}=-\sum_i\frac{Gmm_i}{|\xx -(\ri+\delta\ri)|},
\end{equation}
where $G$ is the gravitational constant.
Expanding the interaction term for small variations of $\delta\ri$ around the  separation distance $|\xx-\ri|$ we obtain
\begin{align}
    H_{\text{int}}=-\sum_iGmm_i&\bigg(\frac{1}{|\xx-\ri|} + \frac{(\xx - \ri)\cdot \delta \ri}{|\xx-\ri|^3} +\dots\bigg).\label{H:int:expand}
\end{align}
The first term is of the form $m\phi({\xx})$, where $\phi$ is the gravitational potential generated by the massive body. It can therefore be absorbed in $H_{\text{m}}$. We assume  a  potential $V({\xx})$ that compensates for the gravitational acceleration, so that  the small particle only performs small oscillations of frequency $\Omega$ around an equilibrium point ${\xx}_0$. Modulo a constant, the Hamiltonian $H_m$ becomes $H_{\text{m}} =\frac{\boldsymbol{p}^2}{2m}+\frac{1}{2}m\Omega^2\delta\xx^2$, where $\delta\xx = {\xx} - {\xx}_0$.

Substituting $\xx = \xx_0 + \delta \xx$ in the second term of  Eq. (\ref{H:int:expand}), we obtain to leading order in $\delta \xx$, 
\begin{eqnarray}
H'_{\text{int}} = - \sum_i \frac{Gmm_i}{|\xx_0-\ri|^3} \bigg(
(\xx_0 - \ri)\cdot \delta \ri + \delta \xx \cdot \delta \ri \nonumber \\  - 3 
\frac{(\xx_0 - \ri)\cdot \delta \ri \, (\xx_0 - \ri)\cdot \delta \xx}{|\xx_0-\ri|^2} \bigg).
\end{eqnarray}
The first term can be absorbed by a shift in the equilibrium positions of the oscillators. The second and third terms give a genuine coupling between  $\delta \xx$ and $\delta \ri$. Assuming that 
the small particle moves along the direction with unit vector ${\bf n}$, we can write $\delta \xx = \delta x {\bf n}$. Then, the coupling term takes the form
\begin{align}
    H''_{\text{int}}=\delta x \sum_i{\bf c}_i \cdot\delta\ri,
\end{align}
where
\begin{equation}\label{coupling:const}
    {\bf c}_i\equiv\frac{Gmm_i}{|\xx_0-\ri|^3} \left[ \frac{3{\bf n}\cdot (\xx_0 - \ri) (\xx_0 - \ri)}{|\xx_0-\ri|^2}  - {\bf n}\right]
\end{equation}
define coupling constants. In this way, the Hamiltonian of the total system can be placed into the form
\begin{equation}\label{tot:Hamiltonian}
    H_{\text{tot}}=H_{\text{m}}+H_{\text{M}}+\delta x \sum_i{\bf c}_i \cdot\delta\ri,
\end{equation}
which is equivalent to the QBM Hamiltonian   that describes the dynamics of a Brownian particle--modelled as a harmonic oscillator--interacting with a thermal environment comprising a large number of independent harmonic oscillators.


\section{Master equation}\label{sec:master}

The dynamics of the reduced density matrix $\rho(\tau)$ of the small particle, obtained by tracing out the degrees of freedom of the environment, can be described by the Hu-Paz-Zhang (HPZ) master equation \cite{HPZ}
\begin{align}\label{HPZ:master}
    \frac{d\rho(\tau)}{d\tau}&=-\frac{i}{\hbar}[\widetilde{H}_m(\tau),\rho(\tau)]-i\gamma(\tau)[\delta x,\{p,\rho(\tau)\}]\nonumber\\
    &\quad -D(\tau)[\delta x,[\delta x,\rho(\tau)]]-f(\tau)[\delta x,[p,\rho(\tau)]].
\end{align}
In the above equation the term $\widetilde{H}_m(\tau)$ represents the particle Hamiltonian with a time-dependent frequency shift $\Omega^2+\delta\Omega^2(\tau)$. The coefficient $\gamma(\tau)$ describes dissipation, while $D(\tau)$ and $f(\tau)$ are diffusion terms. The explicit expressions of the time-dependent coefficients $\delta\Omega^2(\tau), \gamma(\tau), D(\tau)$ and $f(\tau) $, which are rather cumbersome to be reported here, can be found in Refs. \cite{HPZ,HallYu}.  

We note that the HPZ master equation is exact. It is valid for arbitrary temperatures of the environment and  open system-environment coupling strengths. It is derived only under the assumption of a factorized initial state $\rho_{\text{tot}}=\rho\otimes\rho_{\text{env}}$ for the total system, where the environment is in a thermal equilibrium state $\rho_{\text{env}}$ at temperature $T$. 

Straightforward expressions for the coefficients in the master equation \eqref{HPZ:master} can be obtained by employing different approximation schemes during its derivation. Examples include the assumption of a weak coupling between the open system and the environment, and the application of the Markov approximation, which neglects memory effects in the time evolution of the open system \cite{Breuer:open, deVega}. Under the latter two approximations, the diffusion coefficient $D$ is given by
\begin{align}
    D&=\frac{1}{\hbar^2}\int_0^{\infty}ds\,\nu(s)\cos(\Omega s),
\end{align}
where 
\begin{align}\label{noise:kernel}
\nu(\tau):=
\sum_{i}\frac{\hbar |{\bf c}_i|^2}{2m_i\omega_i}\coth\left(\frac{\hbar\omega_i}{2k_BT_i}\right)\cos(\omega_i\tau).
\end{align}
is the so-called {\emph{noise kernel}}. We have assumed a different temperature $T_i$ for each oscillator. Indeed, if the large body is taken to be the Earth, temperature depends on the distance from the center.

In the position representation, $\rho(x,x',\tau)\equiv\bra{x}\rho(\tau)\ket{x'}$, the term of the master equation with  the coefficient $D$ can be expressed as
\begin{align}
    -D[\hat{x},[\hat{x},\rho(\tau)]]\longrightarrow -D(x-x')^2\rho(x,x',\tau),
\end{align}
which indicates that the off-diagonal components ($x\neq x'$) of the reduced density matrix decohere at a rate $D(x-x')^2$ \cite{Zurek,Joos,DecohSchl,Schossh}. Thus, the quantity $D$, which has dimensions [time]$^{-1}\times$[length]$^{-2}$, allows the definition of a \emph{decoherence time} 
\begin{align}
    \tau_{\text{dec}}=\frac{1}{D\Delta x^2},
\end{align}
as the characteristic timescale on which spatial coherences over a distance $\Delta x=x-x'$ becomes suppressed.

\section{The noise kernel}\label{sec:noise}


We treat the heavy body as a solid composed by three-dimensional cubic cells, each shell centered at ${\bf r}_i$, and  representing  a crystal lattice of oscillators (see Fig. \ref{fig:setup}) at constant temperature $T(\ri)$. Each of these cubic cells comprises a total number of normal modes
\begin{align}\label{number:modes}
    N= \frac{1}{3} \int_0^{\infty} g_{\br}(\omega) d\omega,
\end{align}
where  $g_{\br}(\omega)$ is the density of modes whose frequencies lie in the infinitesimal range between $\omega$ to $d\omega$. The index $\br$ denotes spatial dependency into the characterization of the density of modes of each particular cubic cell. Hence, the sum over $i$ in the noise kernel \eqref{noise:kernel}, becomes a sum over all distances $\ri$ and over all modes in a given cell, labelled by $\lambda$. That is, $\sum_i \rightarrow \sum_{\ri} \sum_{\lambda}$.

Employing the continuum limit for the spectrum of the environmental frequencies $\omega_{\lambda}$ in the noise kernel \eqref{noise:kernel} we obtain
\begin{align}
\nu(\tau) =\sum_{\br}\int d\omega&\frac{\hbar G^2m^2m_{\br}}{2\omega}\bigg(|{\boldsymbol{C}}_{\br}|^2g_{\br}(\omega)\nonumber\\&\quad\times\coth\left(\frac{\hbar\omega}{2k_BT(\br)}\right)\cos(\omega\tau)\bigg),
\end{align}
where for later convenience we have written the coupling constants \eqref{coupling:const} as ${\bf c}_i =Gmm_i\boldsymbol{C}_i$.

We will next consider a frequency density of the form  $g_{\br}(\omega) = \alpha_{\br}\omega^{k(\br)}$, up to a cut-off frequency $\omega_c(\br)$, The spectral densities of many systems of interest (e.g., phonon baths, EM field baths) indeed follow a power law. More generally, decoherence is primarily due to the infrared frequency modes of the environment, so it is usually sufficient to consider the dominant behaviour of the spectral density as $\omega \rightarrow 0$; a power-law is generic in this limit \cite{CaLe}. Here, $k>1$ is a constant specific to each cubic cell in the crystal solid. 
The normalization constant $\alpha_{\br}$ is directly computed through Eq. \eqref{number:modes}. It follows that
 \begin{align}\label{density:gen}
    g_{\br}(\omega)=\left\{ \begin{array}{cc} \frac{3N(k(\br)+1)}{\omega_c(\br)^{k(\br)+1}}\omega^{k(\br)}, & \text{for}\quad \omega\leq\omega_c(\br) \\ 0, & \text{for}\quad \omega>\omega_c(\br) \end{array}\right..
\end{align}

Accordingly, the noise kernel is given by
\begin{widetext}
\begin{align}
   \nu(\tau)&= \frac{3}{2} \hbar G^2m^2\sum_{\br}\frac{k(\br) + 1}{\omega_c(\br)^{k(\br)+1}}\int d\omega Nm_{\br}|{\boldsymbol{C}}_{\br}|^2\omega^{k(\br)-1}\coth\left(\frac{\hbar\omega}{2k_BT(\br)}\right)\cos(\omega\tau)\nonumber\\
   &=\frac{3}{2} \hbar G^2m^2\int d^3\br \frac{k(\br) + 1}{\omega_c(\br)^{k(\br)+1}} \rho(\br)|{\boldsymbol{C}}_{\br}|^2\int^{\omega_c}_0 d\omega\,\omega^{k(\br)-1}\coth\left(\frac{\hbar\omega}{2k_BT(\br)}\right)\cos(\omega\tau),
\end{align}
\end{widetext}
where $Nm_{\br}=m(\br)$ is the mass of each cell. In the second equality the continuum-mass limit is taken and   $\rho(\br)$ is the density of the spherical symmetric mass distribution. Hence, to evaluate the noise kernel we need explicit forms for the functions $T(\br), \rho(\br), \omega_c(\br), k(\br)$ in the interior of the compact body.

Here, we will consider the case where these functions are constant. For Earth, this does not affect the order of magnitude of $D$, and hence,  the decoherence time. The density from Earth's center to the surface changes at most by a factor of 5; the temperature is significantly higher at the center, but the contribution of internal layers to $D$ is strongly suppressed. 

For constant functions $T(\br), \rho(\br), \omega_c(\br), k(\br)$, and for a spherical body of radius $R$, we only need to compute the integral
\begin{align}
    I(x_0,R)=\int_{r\leq R}  d^3 \br|{\boldsymbol{C}}_{\br}|^2. \label{ixr}
\end{align}
For $x_0$ along the z-axis,  particle motion along  $\boldsymbol{n}=(0,1,0)$, and $\br=(r\sin\theta\cos\phi,r\sin\theta\sin\phi,r\cos\theta)$, we obtain
\begin{widetext}
\begin{align}\label{integral:tot}
    I(x_0,R)=\int_{r\leq R} \frac{d^3\br }{|\xx_0-\br|^6}\bigg(1+\frac{9r^4\sin^4\theta\sin^2\phi}{|\xx_0-\br|^4}-\frac{6r^2\sin^2\theta\sin^2\phi}{|\xx_0-\br|^2}+\frac{9r^2\sin^2\theta\sin^2\phi}{|\xx_0-\br|^4}(x_0-r\cos\theta)^2\bigg).
\end{align}
\end{widetext}
We evaluate this integral in Appendix \ref{appendix}. 
We note that for a body at distance $d << R$ from the surface of the Earth, $x_0 = R + d$, and 
\bey
I(x_0, R) \simeq \frac{3\pi}{16 d^3}.
\eey

The noise kernel reads
\begin{align}
   \nu(\tau)&=\frac{9(k+1)\hbar G^2m^2M}{8\pi\omega_c^{k+1}R^3}I(x_0,R)\nonumber\\
   &\quad \times\int^{\omega_c}_0 d\omega\,\omega^{k-1}\coth\left(\frac{\hbar\omega}{2k_BT}\right)\cos(\omega\tau),
\end{align}
where $M$ is the total mass of the massive body.

 The noise kernel diverges for $d \rightarrow 0$. This means that the decoherence rate is very sensitive on the outer layers of the massive body. When using the noise kernel to model a specific experiment, we must split the mass distribution into a spherical part that corresponds to the bulk of the Earth, and a part that models the immediate surroundings of the particle in the experiment. In this sense, $d$ is best understood as an effective distance of the particle from the surrounding masses. For particles near the surface of the Earth, the effective $d$ is of the order of the length-scales that characterize the laboratory.

\subsection{The decoherence time}

\begin{table*}[t!]
\begin{center}
\begin{tabular}{ |c|c|c|c| } 
 \hline
Massive body  & Distance from surface & Mass $m$ & Decoherence time \\ 
  \hline\hline
 Earth &  $10cm$ & $100kg$ &  $\tau_{\text{dec}}\sim 1s$ \\ 
 Earth &  $10cm$ & $10^5kg$ &   $\tau_{\text{dec}}\sim 10^{-6}s$\\ 
 Earth &    $400km$& $100kg$&      $\tau_{\text{dec}}\sim 10^{20}s$   \\
 Earth &    $10cm$ & $10^{-27}kg$&      $\tau_{\text{dec}}\sim 10^{58}s$\\
 White dwarf & $10cm$ & $100kg$ & $\tau_{\text{dec}}\sim 10^{-6}s$ \\
 White dwarf & $10m$ & $100kg$ & $\tau_{\text{dec}}\sim 1s$ \\
 \hline
\end{tabular}
\end{center}
\caption{The decoherence time in different regions of the parameter space. The distance of $400km$ corresponds to the average altitude at which the ISS maintains its orbit. The mass $m$ of $10^{-27} kg$ corresponds to the atomic mass. 
}\label{table}
\end{table*}
We next focus our attention to the case of a function of the density of states that is quadratic to frequency, i.e, we choose $k=2$ in Eq. \eqref{density:gen}. This is equivalent to the Debye model for the density of states of solids \cite{Ashcroft, Pathria}--see, also, Appendix \ref{appendix:Debye}.  The cut-off frequency $\omega_c$ coincides with the  Debye frequency $\omega_D$ of the solid, and the environmental degrees of freedom are characterised by an Ohmic spectral density (in an Ohmic environment, the damping force experienced by the Brownian particle is linear to its velocity \cite{HPZ, Weiss}).

The decoherence rate reads
\begin{align}
    D&=\frac{27G^2m^2M}{8\pi\hbar\pi\omega_D^{3}R^3}I(x_0,R)\nonumber\\&\times\int_0^{\infty}d\tau\cos(\Omega\tau)\int^{\omega_D}_0 d\omega\,\omega\coth\left(\frac{\hbar\omega}{2k_BT}\right)\cos(\omega\tau).
\end{align}
At the limit $\omega_D\to\infty$, we obtain
\begin{align}
    D=\frac{27G^2m^2M}{8\pi\hbar\omega_D^{3}R^3}I(x_0,R)\left(\frac{\pi\Omega}{2}\coth\left(\frac{\hbar\Omega}{2k_BT}\right)\right).
\end{align}

In the physically relevant regime, $k_BT\gg\hbar\Omega$, we  approximate $\coth(\hbar\Omega/2k_BT)\approx 2k_B T/\hbar\Omega$. Then, 
\begin{align}\label{decoh:D}
    D=\frac{27G^2m^2M}{8R^3}\frac{k_B T}{\hbar^2\omega_D^3}I(x_0,R).
\end{align}

In Table \ref{table}, we present the decoherence time for a superposition of two localized states with mean separation of $\Delta \xx = 1m$ in various cases, calculated by means of \eqref{decoh:D}. For a human on a distance $10cm$ from the surface of Earth ($R_{\oplus}=6.4\times10^6m, M_{\oplus}=5.9\times10^{24}kg$) we find a decoherence time of order of $1s$. Earth is described as a Debye solid with a Debye frequency of order $\omega_D\sim 10^{13}Hz$. We also give the decoherence rate for a white dwarf. 
During the crystallization phase of its ion lattice, a white dwarf ($M\sim 1M_{\odot}=2\times10^{30}kg$, $R\sim R_{\oplus}$) can also be described as a Debye solid \cite{Shapiro}. In this case, the Debye frequency is of order $\omega_D\sim10^{18}Hz$.

\section{Conclusions}\label{sec:concl}
We showed that any quantum particle within the gravitational field of a massive body is subject not only to the gravitational pull, but also to non-unitary dynamics that originate from the intrinsic fluctuations of the matter in the massive body, and mediated by the gravitational field.

The non-unitary contribution is of second-order to the gravitational constant, and for this reason, it is implicitly assumed to be negligible, and it is usually ignored. However, the strength of a non-unitary term depends on the strength of the fluctuations, and may become quite strong, for example, at high temperatures, or in an environment with a strong concentration of modes in the deep infrared. In any case, the gravitational field will mediate a decoherence process, from which no experiment in the gravitational field of the massive body can be shielded. 

Our analysis demonstrated that this effect  does not affect the regime of interest to near-future experiments with macroscopic quantum systems. It is expected to appear for superpositions at the human scale. 

We note that our treatment is completely general: it can be applied to describe quantum phenomena within the gravitational field of any compact body that is compatible with the Newtonian description of gravity. It allows for the incorporation of fine details, such as density and temperature gradients, or the gravitational influence of the immediate surroundings of the quantum system.

Our analysis in this paper is restricted to positional decoherence. However, this is not the only way that the environment can destroy quantum coherences. Since the environmental degrees of freedom are entangled with those of the quantum particles, it will also affect the generation of entanglement in multi-partite systems due to the gravitational force. It is therefore important to analyze the degree that proposed experiments for gravity-induced entanglement are affected by the presence of gravity-mediated decoherence.

Finally, we note that our analysis straightforwardly applies to the decoherence induced by a large charged body on microscopic charged particles. In this case, the Coulomb-mediated decoherence effects are much stronger, and hence, experimentally accessible.



\hspace{2cm}


\onecolumngrid

\appendix

\section{Explicit expressions for the diffusion constant}\label{appendix}
We evaluate the quantity $I(x_0, R)$ defined by Eq. (\ref{ixr}). Assuming that  
$\xx=(0,0,x_0)$, $\boldsymbol{n}=(0,1,0)$ and $\br=(r\sin\theta\cos\phi,r\sin\theta\sin\phi,r\cos\theta)$, 
\begin{align}\label{integral:tot}
    I(x_0, R) =\int_{r\leq R} \frac{d^3\br }{|\xx_0-\br|^6}\bigg(1+\frac{9r^4\sin^4\theta\sin^2\phi}{|\xx_0-\br|^4}-\frac{6r^2\sin^2\theta\sin^2\phi}{|\xx_0-\br|^2}+\frac{9r^2\sin^2\theta\sin^2\phi}{|\xx_0-\br|^4}(x_0-r\cos\theta)^2\bigg).
\end{align}
For a sphere of radius $R$, we define ${\bf z} = {\bf x}_0/R$, ${\bf x} =  {\bf r}/R$, to obtain
\bey
I(x_0, R) = \frac{1}{R^3} F(z),
\eey
where 
\bey
F(z) = 2 \pi\int_0^1 dx \int_{-1}^1 d\xi  \frac{x^2}{|{\bf z} - {\bf x}|^6} \left( 1 + \frac{9x^4(1 - \xi^2)}{2|{\bf z} - {\bf x}|^4} - \frac{3x^2(1 - \xi^2)}{|{\bf z} - {\bf x}|^2} + \frac{9z^2x^2(1 - \xi^2)}{2|{\bf z} - {\bf x}|^4} - \frac{9zx^3(1 - \xi^2)\xi}{|{\bf z} - {\bf x}|^4}\right)
\eey
and we wrote $\xi = {\bf z}\cdot {\bf x}/(zx)$.

We carry out the integration over $\xi$, to obtain

\bey
F(z) &=& \frac{4\pi}{3} \frac{1}{(z^2-1)^3} + 4\pi \int_0^1 dx \frac{x^4}{(x^2 - z^2)^4} \nonumber \\
&=& \frac{4\pi}{3} \frac{1}{(z^2-1)^3} + \pi \frac{3z+ 8z^3-3z^5}{12 z^3(z^2-1)^3} + \frac{\pi}{4 z^3} \mbox{arccoth}(z).
\eey
For $z = 1 + x$, with $x << 1$, we obtain
\bey
F(x) \simeq \frac{3\pi}{16x^3}.
\eey

\section{Debye solid}\label{appendix:Debye}
A solid can be viewed as an ordered array of atoms, where each atom is fixed to a lattice site and can oscillate about its equilibrium position. The Debye model \cite{Ashcroft, Pathria} treats these atomic vibrations as sound waves that propagate through the crystal lattice at the speed of sound $\upsilon_s$. The vibrational energy is quantized, and the quanta of vibrational energy are known as \emph{phonons}, analogous to photons in electromagnetic waves. The frequencies of these phonons are linearly related to their wave vectors $\boldsymbol{k}$ through the dispersion relation $\omega(\boldsymbol{k})=\upsilon_s|\boldsymbol{k}|$.

The Debye model allows for a continuum spectrum of oscillation frequencies, resulting in a total number of normal modes of vibration equal to $3N$,  expressed as
\begin{align}
    3N=\int_0^{\omega_D}g(\omega)d\omega,
\end{align}
where $N$ is the number of atoms in the crystal, $g(\omega)$ is the density of normal modes meaning  that $g(\omega)d\omega$ represents the  number of normal modes with frequencies in the infinitesimal range between $\omega$ and $\omega+d\omega$, and $\omega_D$ is a maximum allowed phonon frequency, known as the {\emph{Debye frequency}}. The Debye frequency acts as a cut-off for phonon frequencies and is determined by the minimum wavelength allowed for sound wave propagation in the lattice, constrained by the finite interatomic distance.

The Debye model approximates the density of modes as
\begin{align}
g(\omega) = \begin{cases}
\frac{9N}{\omega_D^3}\omega^2,& \text{for } \omega \leq \omega_D \\
0, & \text{for } \omega > \omega_D
\end{cases}.
\end{align}
The Debye frequency is defined by
\begin{align}
    \omega_D^3=18\pi^2n\left(\frac{1}{\upsilon_L^3}+\frac{2}{\upsilon_T^3}\right)^{-1}
\end{align}
where $n$ is the atomic density, $\upsilon_L$ is the velocity of the longitudinal sound mode, and $\upsilon_T$ is the velocity of the transverse sound modes.

\twocolumngrid

\end{document}